\documentclass[10pt]{aastex}

\usepackage{amsmath,natbib}
\usepackage{rotating}
\usepackage{emulateapj5}
\usepackage{graphicx}

\def\spose#1{\hbox to 0pt{#1\hss}}
\def\lta{\mathrel{\spose{\lower 3pt\hbox{$\mathchar"218$}}
     \raise 2.0pt\hbox{$\mathchar"13C$}}}
\def\gta{\mathrel{\spose{\lower 3pt\hbox{$\mathchar"218$}}
     \raise 2.0pt\hbox{$\mathchar"13E$}}}

\newcommand{\hii}{\text{H}{\sc ii}} 

\shorttitle{Anomalous microwave emission from RCW175.}
\shortauthors{C.~Dickinson et al.}

\begin{document}

\title{Anomalous microwave emission from the \hii~region RCW175}

\author{
C.~Dickinson,\!\altaffilmark{1}
R.~D.~Davies,\!\altaffilmark{2}
J.~R.~Allison,\!\altaffilmark{3}
J.~R.~Bond,\!\altaffilmark{4}
S.~Casassus,\!\altaffilmark{5}
K.~Cleary,\!\altaffilmark{6}
R.~J.~Davis,\!\altaffilmark{2}
M.~E.~Jones,\!\altaffilmark{3}
B.~S.~Mason,\!\altaffilmark{7}
S.~T.~Myers,\!\altaffilmark{8}
T.~J.~Pearson,\!\altaffilmark{6}
A.~C.~S.~Readhead,\!\altaffilmark{6}
J.~L.~Sievers,\!\altaffilmark{4}
A.~C.~Taylor,\!\altaffilmark{3}
M.~Todorovi\'c,\!\altaffilmark{2}
G.~J.~White,\!\altaffilmark{9}
P.~N.~Wilkinson\!\altaffilmark{2}
} 

\altaffiltext{1}{Infrared Processing and Analysis Center, California Institute of Technology, M/S 220-6, 1200 E. California Blvd., Pasadena, CA 91125, U.S.A.}
\altaffiltext{2}{Jodrell Bank Observatory, University of Manchester, Lower Withington, Macclesfield, Cheshire, SK11 9DL, U.K.}
\altaffiltext{3}{Oxford Astrophysics, University of Oxford, Denys Wilkinson Building, Keble Road, Oxford, OX1 3RH, U.K.}
\altaffiltext{4}{Canadian Institute for Theoretical Astrophysics, University of Toronto, Toronto, Canada.} 
\altaffiltext{5}{Departamento de Astronom{\'\i}a, Universidad de
Chile, Casilla 36-D, Santiago, Chile.}  \altaffiltext{6}{Chajnantor
Observatory, California Institute of Technology, M/S 105-24, Pasadena,
CA 91125, U.S.A.}  \altaffiltext{7}{National Radio Astronomy
Observatory, Green Bank, WV.}  \altaffiltext{8}{National Radio
Astronomy Observatory, Socorro, NM.}  \altaffiltext{9}{Rutherford
Appleton Laboratory, Didcot, OX11 0QX, U.K.}


\begin{abstract}
We present evidence for anomalous microwave emission in the RCW175
\hii~region. Motivated by 33~GHz $13\arcmin$ resolution data from the Very Small Array (VSA),
we observed RCW175 at 31~GHz with the Cosmic Background Imager (CBI)
at a resolution of $4\arcmin$. The region consists
of two distinct components, G29.0-0.6 and G29.1-0.7, which are
detected at high signal-to-noise ratio. The integrated flux density is
$5.97\pm0.30$~Jy at 31~GHz, in good agreement with the VSA. The 31~GHz
flux density is $3.28\pm0.38$~Jy ($8.6\sigma$) above the expected value from optically thin free-free emission
based on lower frequency radio data and thermal dust constrained by
IRAS and WMAP data. Conventional emission mechanisms such as optically
thick emission from ultracompact \hii~regions cannot easily account
for this excess. We interpret the excess as evidence for electric dipole
emission from small spinning dust grains, which does provide an
adequate fit to the data.
\end{abstract}




\section{Introduction}

In recent years there has been mounting observational evidence for a
new diffuse component emitting at frequencies $\approx
10-60$~GHz. The anomalous microwave emission was first detected at
14 and 32~GHz by \cite{Leitch97}. Since then, a similar picture has
emerged both at high latitudes
\citep{Banday03,deOliveira-Costa04,Fernandez-Cerezo06,Hildebrandt07,Bonaldi07}
and from individual Galactic sources
\citep{Casassus04,Casassus06,Casassus07,Watson05,Scaife07,Dickinson07},
although negative detections have also been reported
\citep{Dickinson06,Scaife08}.  The spectral index between 20 and 40~GHz
is $\alpha \approx -1.1$ \citep{Davies06} with some evidence of
flattening at $\sim10-15$~GHz
\citep{Leitch97,deOliveira-Costa04,Hildebrandt07}. The emission
appears to be very closely correlated with far-IR data suggesting a
dust origin. Various emission mechanisms have been suggested,
including hot ($T\sim10^{6}$~K) free-free \citep{Leitch97}, flat
spectrum synchrotron \citep{Bennett03b}, spinning dust
\citep{Draine98a,Draine98b} and magnetic dust \citep{Draine99}. The
overall picture is still very unclear and new data covering the range
$10-60$~GHz are urgently needed.

RCW175 \citep{Rodgers60} is a diffuse \hii~region, which consists of a
``medium brightness'' optical filament (G29.1-0.7, S65) $\sim 7\arcmin
\times 5\arcmin$ in extent, and a nearby compact source (G29.0-0.6),
which is heavily obscured by dust. Although the filament is clearly
seen in high resolution data, the compact counterpart is considerably
brighter. The ionization is thought to be provided by a single B-1 II
type star, which forms part of a 5-star cluster
\citep{Forbes89,Sharpless59} at a distance of 3.6~kpc. 

Observations made with the Very Small Array (VSA) at 33~GHz
\citep{Watson03,Dickinson04} as part of a Galactic plane survey
(Todorovi\'c et al., in prep.) indicate that RCW175 is anomalously
bright by a factor of $\approx 2$, when compared with lower frequency
data. In this {\it Letter}, we present accurate Cosmic Background
Imager (CBI) $31$~GHz observations of RCW175 and make a comparison
with ancillary radio/FIR data. We find that the emission at $31$~GHz
is significantly above what is expected from a simple model of
free-free and vibrational dust emissions. 


\section{Data}
\label{sec:data}

\subsection{31~GHz: Cosmic Background Imager}
\label{sec:cbi}

The CBI is a $26-36$~GHz
13-element comounted interferometer, operated at the Chajnantor
Observatory, Chile. The original CBI used 0.9~m dishes to provide high
temperature sensitivity measurements on angular scales $\sim
30\arcmin-6\arcmin$ \citep{Padin02,Readhead04a}. Recently, it has been
upgraded with 1.4~m dishes (CBI2) to give increased temperature
sensitivity on angular scales $\approx 4\arcmin-15\arcmin$ (Taylor et al., in prep.).

We observed RCW175 at RA$=18^{\rm h}46^{\rm m}40^{\rm s}$,
Dec$=-03^{\rm d}46^{\rm m}00^{\rm s}$ (J2000), on 4 nights in May
2007, with a total integration time of 6~hr with CBI2. Boresight rotations of
the array were used to improve the $u,v$-coverage. Ground spillover
was removed by subtracting observations of a comparison field, 8~min later in RA, observed at the same hour angles. Jupiter was the primary amplitude/phase calibrator with
absolute calibration tied to a Jupiter temperature of $146.6\pm0.75$~K
at 33~GHz \citep{Hill08}. Observations of secondary calibrators
showed that the pointing was good to better than $1\arcmin$.

\begin{figure*}[!h]
\centering \includegraphics[width=1.0\textwidth, angle=0]{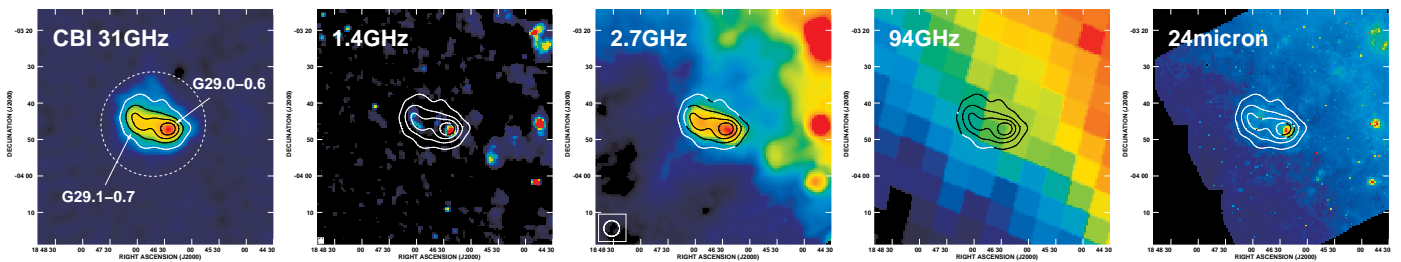}
\caption{Multi-frequency maps of the RCW175 region. From left to right: CBI 31~GHz CLEANed map, NVSS 1.4~GHz, Effelsberg 2.7~GHz, WMAP 94~GHz and {\it Spitzer} MIPS $24~\mu$m. Angular resolutions and references are given in Table~\ref{tab:int_flux}. Contours are CBI 31~GHz at $20,40,60,80\%$ of the peak brightness, 1.04~Jy/beam. The CBI primary beam (FWHM) is shown as a dashed line. }
\label{fig:multi_maps}
\end{figure*}

A uniform-weighted, CLEANed map is shown in
Fig.~\ref{fig:multi_maps}. The synthesized beam is
$4\arcmin.3\times4\arcmin.0$ and the primary beam is approximately
Gaussian with FWHM $28\arcmin.2~(30{\rm GHz}/\nu)$. Corrections for
the primary beam were made directly to the CLEAN components at each
frequency; the bulk of the emission fits well within the extent of the
primary beam. The total flux density in the map is 6.4~Jy with a peak
brightness of 1.04~Jy/beam and the noise level is 27~mJy/beam.

The CBI map detects and resolves both components in the RCW175 region. G29.0-0.6 in the west is more compact and is significantly
brighter than the more diffuse G29.1-0.7 in the east. Although there
is some extended emission almost all ($93\%$) of the flux can be fitted
by two Gaussian components. Note that the entire extent
of RCW175 is equivalent to $\approx 13$ beam areas, thus the maximum
CLEAN bias is $\approx 0.35$~Jy. We fitted two elliptical Gaussians plus
a baseline offset using the {\sc aips} task {\sc jmfit}. At 31~GHz we
find that G29.0-0.6 has a deconvolved size $5\arcmin.4\times5\arcmin.1$ and integrated flux
density, $S_i=2.20~$Jy. G29.1-0.7 is $10\arcmin.0\times8\arcmin.2$ and $S_i=3.76$~Jy. Simulations
showed that CBI flux-loss due to the limited $u,v$-coverage is $\lesssim 10\%$ when fitting two Gaussians to the bulk of the emission.

\subsection{Ancillary radio/FIR data}
\label{sec:other_data}

Table~\ref{tab:int_flux} lists frequencies, angular resolutions and references for the data\footnote{Data were
downloaded from the {\it Skyview} website
(http://skyview.gsfc.nasa.gov), the MPIfR Image Survey Sampler website
(http://www.mpifr-bonn.mpg.de/survey.html), the LAMBDA website
(http://lambda.gsfc.nasa.gov/) and the IRSA website
(http://irsa.ipac.caltech.edu)} used in this {\it Letter} and Fig.~\ref{fig:multi_maps} shows selected maps centered on RCW175. The
radio maps from 1.4~GHz to 14.35~GHz show enhanced emission at the
same location as that seen in the CBI image with the bright, more
compact, hotspot in the west (G29.0-0.6) and the fainter component
to the east (G29.1-0.7).  For data with resolutions $\sim 4\arcmin$, the sub-components
are not well separated and the bulk of the emission is confined to a
region $\sim 10\arcmin \times 5\arcmin$. At these resolutions the
region looks like one extended object with G29.0-0.6 dominating at
one end. The higher resolution NVSS 1.4~GHz/{\it Spitzer}~$24~\mu$m
data show  the more compact G29.0-0.6
has an extent of $\approx 2\arcmin$. A filament of
emission to the east (G29.1-0.7) is coincident with the filament
seen in high resolution optical/IR images. There is low-level emission in
the region in between, which is particularly evident in the $24~\mu$m
image.

WMAP 5-yr total-intensity maps \citep{Hinshaw08} covering 23-94~GHz show no significant detection of
emission above the Galactic background at the location of RCW175. The 94~GHz map, at $\approx 12\arcmin.6$ resolution, is however
useful for placing an upper limit on the thermal dust contribution (see
\S\ref{sec:int_flux}).


\section{Flux density spectrum}
\label{sec:analysis}

\subsection{Integrated flux density spectrum}
\label{sec:int_flux}

The integrated flux density in each map was calculated both by fitting
multiple Gaussians (with baseline offset) and by integrating over a
given aperture. For the interferometric data, both methods gave
roughly consistent results within the errors. For total-power data,
the aperture value could be over-estimated due to the background
level. Moreover, a possible offset of $\sim 0.\arcmin5$ in the CBI map
could result in a bias depending on the exact aperture location; an offset of $\approx 0\arcmin.5$ does appear to be visible when comparing CBI contours with the NVSS 1.4~GHz map (Fig.~\ref{fig:multi_maps}). For
these reasons, we chose the Gaussian fitting method as described in \S\ref{sec:cbi}. 

Integrated flux densities for RCW175 and the G29.0-0.6 component are given in
Table~\ref{tab:int_flux} and plotted in Fig.~\ref{fig:spec}. Absolute
calibration errors (assumed to be 10\%  if not known) were added in
quadrature to the fitting error reported by {\sc jmfit}. We include
the 33~GHz value for RCW175 from the VSA, which is in good agreement with
the CBI value. The lower resolution ($\approx 13\arcmin$) VSA data
meant that a single Gaussian fit was adequate and that flux loss was
negligible. Since there is a correction factor of 1.53 for the primary
beam at the position of RCW175, we assigned a conservative 10\% error
to account for any small deviations from the assumed Gaussian primary
beam. An upper limit was inferred from the WMAP W-band (94~GHz) map by
fitting for a parabolic baseline in the vicinity of RCW175 to account
for the Galactic background, and calculating the $3\sigma$ of the
residual map.

\subsection{SED fitting}
\label{sec:sed_fitting}

For a classical \hii~region, the radio/microwave spectrum consists of
free-free emission at radio wavelengths and vibrational dust emission at
infrared wavelengths. At the lowest radio frequencies (typically
$\lesssim 1$~GHz), the free-free emission is optically thick and follows a
$\alpha\approx +2$ spectrum\footnote{Throughout this {\it Letter}, we use the spectral index convention for flux density given by $S \propto \nu^{\alpha}$.}. Above the turnover frequency (typically
$\sim 1$~GHz), the emission becomes optically thin and follows a
$\alpha\approx -0.1$ power-law, with very little dependence on
physical conditions \citep{Rybicki79,Dickinson03}. At high frequencies
(typically $>100$~GHz), vibrational dust emission becomes
dominant. Vibrational dust is well-represented by a modified
black-body function, $\nu^{\beta+2} B(\nu,T_{\rm dust})$, where $\beta$
is the dust emissivity index and $B(\nu,T_{\rm dust})$ is the black-body
function for a given dust temperature, $T_{\rm dust}$.

\begin{figure*}[!h]
\centering \includegraphics[width=0.4\textwidth, angle=0]{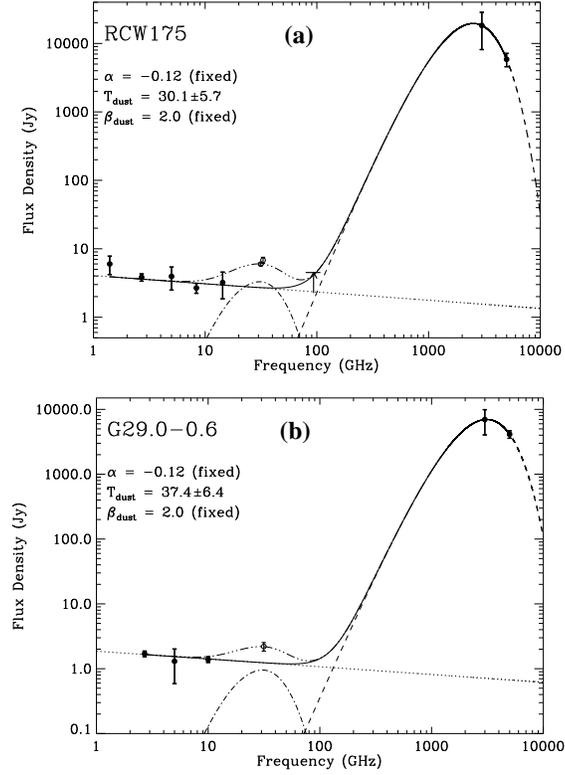}
\caption{Integrated flux density spectrum for a) both components combined (RCW175) and b) the brighter component (G29.0-0.6). Filled circles represent data fitted by a power-law with fixed spectral index (dotted line) and a modified black-body with fixed emissivity (dashed line). The \cite{Draine98b} CNM spinning dust spectrum (dot-dashed line) has been fitted to the 31/33~GHz data.}
\label{fig:spec}
\end{figure*}

The SED for RCW175 and the G29.0-0.6 component is shown in Fig.~\ref{fig:spec}. The low frequency ($<15$~GHz) data
were fitted by an optically thin free-free power-law spectrum ($S\propto \nu^{\alpha}$) fixed to
the theoretical value, $\alpha=-0.12$ for $T_{e}\approx 8000$~K and
frequencies around $\sim 10$~GHz. When fitting for the index we found
$\alpha=-0.32\pm0.15$, which is consistent with this value. The infrared data ($60/100~\mu$m) were fitted by a
single temperature dust component using a simple modified black-body
curve, $S \propto \nu^{\beta+2}B(\nu,T_{\rm dust})$. We did not attempt to fit data at shorter wavelengths. Since there are
no data points in the sub-mm ($\sim 100-1000$~GHz), the emissivity index was fixed at $\beta=+2.0$, which
is typical for \hii~regions (e.g. \citet{Gordon88}). Flatter indices
in the range $\beta \approx 1-2$ have been observed while the
preferred value for $T_{\rm dust}\approx 30$~K is $\beta \approx +1.6$
\citep{Dupac03}. However, the 94~GHz upper limit does not allow
flatter indices than $\beta \approx 2$. The best-fitting dust temperatures were $T_{\rm dust}=30.1\pm5.7$~K and $T_{\rm dust}=37.6\pm6.4$~K, for RCW175 and G29.0-0.6, respectively.

 From Fig.~\ref{fig:spec} it is clear that the $\sim 30$~GHz flux
 densities are significantly higher than the simple model can allow
 for. This model predicts a 31~GHz flux density for RCW175 of $2.69\pm
 0.24$~Jy, compared to the measured value of $5.97\pm0.30$~Jy. This
 corresponds to an excess over the free-free emission of
 $3.28\pm0.38$~Jy ($8.6\sigma$). Adopting the best-fit spectral index at low frequencies
 results in a larger excess at 31~GHz, although at a reduced significance level
 ($6.4\sigma$) due to the increased error when fitting for an
 additional parameter. The brighter component of G29.0-0.6 shows a similar level of excess (Fig.~\ref{fig:spec}).


\section{Discussion}
\label{sec:discussion}

A peaked (convex) spectrum is required to produce excess emission at $\sim 30$~GHz without exceeding the upper limit at 94~GHz. The $3\sigma$ upper limit at 94~GHz strongly
rules out a cold thermal dust component, or a flatter dust
emissivity.  Convex
spectra can be produced by optically thick ultracompact (UC)
\hii~regions, gigahertz-peaked spectrum (GPS) sources, or some new
mechanism such as electro-dipole and magneto-dipole radiation.

UC \hii~regions are self-absorbed up to higher frequencies because of
extremely high densities. Indeed it is possible to fit the radio data including the
VSA/CBI data by including a homogenous compact \hii~region with
angular size $\sim 1\arcsec$ and emission measure of $\sim
3\times10^{9}$~cm$^{-6}$pc. Such parameters are within observed limits
\citep{Wood89}. However, the additional component produces too much
flux ($\gtrsim 10$~Jy) at 94~GHz. For the same reasons, magneto-dipole radiation can also be excluded as a major contributor.

GPS sources are high redshift radio sources in which the
radio jets have been highly confined and the synchrotron emission is
self-absorbed. A search through the NVSS catalogue \citep{Condon98} revealed no bright
($\gtrsim 100$~mJy) compact radio sources in this region and no bright ($\gtrsim 20$~mJy), point-like sources appear in the NVSS 1.4~GHz map. 

The anomalous emission could be due to electric dipole radiation from small rapidly
spinning dust grains. If such grains have a residual dipole moment and
spin at GHz frequencies, they will emit over a narrow range of
frequencies. Fig.~\ref{fig:spec} shows a
typical spinning dust spectrum using the model of \cite{Draine98b} for
the Cold Neutral Medium (CNM), scaled to fit the VSA/CBI
data. The highly peaked spectrum allows an adequate fit to the data
both at $\sim 10$~GHz and yet consistent with the 94~GHz upper limit. We note that the appropriate
model for Warm Ionized Medium (WIM) peaks at a slightly lower
frequency and predicts too much flux at $\sim 10$~GHz. However, the
models are fairly generic and allow considerable freedom in the dust
parameters \citep{Finkbeiner04a}. A similar situation is observed in
the \hii~region G159.6-18.5 \citep{Watson05} and the dark cloud LDN1622
\citep{Casassus06}.  

The \cite{Draine98b} models are expressed in terms of the intensity per hydrogen column density, in units of
Jy~sr$^{-1}$~cm$^2$ per H atom. We estimate an average column density
of hydrogen for RCW175 $N$(H)$\approx 4.4\times10^{22}$~cm$^{-22}$ using the
canonical factor $2.13\times10^{24}$~H~cm$^{-2}$ per unit of $\tau_{100\mu{\rm
m}}$ \citep{Finkbeiner04b}. The best-fitting spinning dust model in Fig.~\ref{fig:spec}
corresponds to $N$(H)$=4.4\times10^{22}$~cm$^{-22}$. The observed
levels are therefore consistent with the emissivities of
\cite{Draine98b}. It is also remarkable that the amplitude of the anomalous
component is similar to the emissivity relative to $100~\mu$m
observed at high Galactic latitudes \citep{Davies06}, which is
equivalent to $1$~Jy at 30~GHz per 6000~Jy at 100~$\mu$m. This does not appear to be the case for all \hii~regions, which have so far shown to have a reduced radio-to-dust emissivity \citep{Dickinson07,Scaife07}. High resolution, multi-frequency data in the range $10-100$~GHz are needed to confirm this result and to investigate the nature of anomalous emission.


\section{Conclusions}
\label{sec:conclusions}

Using data from the VSA and CBI, we have observed excess emission at
$\sim 30$~GHz from the \hii~region RCW175. The flux density spectrum
indicates that about half of the flux in this region is from optically
thin free-free emission leaving about half unaccounted for. An upper
limit at 94~GHz from WMAP data constrains the contribution of thermal
dust emission. We have discarded optically thick free-free emission from
UC~\hii~regions and GPS sources as possible candidates for this excess; an upper limit at 94~GHz
and high resolution radio data rule these out as the primary
contributor at $\sim 30$~GHz. We interpret the excess as electric
dipole radiation from small rapidly spinning dust grains as predicted
by \cite{Draine98b}. These models provide a reasonable fit to the data
that is consistent both in terms of spectral shape and
emissivity. High resolution, multi-frequency data in the range
$10-100$~GHz are needed to confirm this result and to investigate the
nature of anomalous emission.


\acknowledgements{This work was supported by the Strategic Alliance for the Implementation of New Technologies (SAINT - see www.astro.caltech.edu/chajnantor/saint/index.html) and we are most grateful to the SAINT partners for their strong support. We gratefully acknowledge support from the Kavli
  Operating Institute and thank B. Rawn and S. Rawn Jr. The CBI was supported by NSF grants 9802989, 0098734
  and 0206416, and a Royal Society Small Research Grant. We are
  particularly indebted to the engineers who maintain and operate the
  CBI: C. Achermann, R. Bustos, C. Jara, N. Oyarace, R. Reeves, M. Shepherd and C. Verdugo. CD thanks Roberta
  Paladini and Bill Reach.  CD acknowledges support from the U.S. {\it Planck}
  project, which is funded by the NASA Science Mission Directorate. SC
  acknowledges support from FONDECYT grant 1030805, and from the
  Chilean Center for Astrophysics FONDAP 15010003. AT acknowledges support from Royal Society and STFC research fellowships.}

\bibliographystyle{apj}
\bibliography{refs.bib}


\begin{table}
\begin{center}
\caption{Flux densities for RCW175 and G29.0-0.6. }
\begin{tabular}{cccccc} 
\tableline \tableline
Frequency     &Telescope/    &Angular resolution &Reference         &Flux density   &Flux density \\
(GHz)         &survey        &(arcmin)           &for data          &RCW175 (Jy)    &G29.0-0.6 (Jy)      \\
\tableline
1.4           &Green Bank 300 ft     &$9.4\times10.4$   &\cite{Altenhoff70}   &$6.0\pm1.8$           &Not resolved \\
2.7           &Effelsberg 100~m &4.3                &\cite{Reich90}    &$3.83\pm0.38$         &$1.70\pm0.17$ \\
5             &Parkes 64~m      &4.1                &\cite{Haynes78}   &$3.90\pm0.39$         &$1.30\pm0.71$ \\
8.35          &Green Bank 13.7 m  &9.7                &\cite{Langston00} &$2.68\pm0.35$         &Not resolved \\
10            &Nobeyama 45~m      &3                  &\cite{Handa87}    &Not reported                      &$1.39\pm0.14$ \\
14.35         &Green Bank 13.7 m  &6.6                &\cite{Langston00} &$3.21\pm1.83$         &Not resolved \\
31            &CBI                &$4.3\times4.0$     &This work         &$5.97\pm0.30$         &$2.20\pm0.33$ \\
33            &VSA                &$13.1\times 10.0$  &Todorovic et al. (in prep.) &$6.83\pm0.68$ &Not resolved \\
94            &WMAP               &$\approx 12.6$     &\cite{Hinshaw08}  &$<4.5~(3\sigma)$      &$<4.5~(3\sigma)$  \\
2997 ($100~\mu$m) &IRAS           &$\approx 4$    &\cite{Beichman88}&$18320\pm10190$        &$6970\pm2930$ \\
4995 ($60~\mu$m)  &IRAS           &$\approx 4$    &\cite{Beichman88} &$5864\pm1292$         &$4131\pm548$ \\
12875 ($24~\mu$m) &{\it Spitzer} MIPSGAL &$0.033$        &http://irsa.ipac.caltech.edu/               &Not fitted            &Not fitted \\
\tableline
\end{tabular}
\label{tab:int_flux}
\end{center}
\end{table}


\end{document}